\documentstyle[aps,epsfig]{revtex}

\widetext

\begin{document}
\title{Heating and decoherence suppression using decoupling techniques}
\author{D. Vitali and P. Tombesi}
\address {Dipartimento di Matematica e Fisica, and Unit\`a INFM,
Universit\`a di
Camerino \\ via Madonna delle Carceri, I-62032 Camerino, Italy}
\date{\today}
\maketitle

\begin{abstract}
We study the application of decoupling techniques to the case of a 
damped vibrational mode of a chain of trapped ions, which can be used as 
a {\em quantum bus} in linear ion trap quantum computers. We show that
vibrational heating could be efficiently suppressed using 
appropriate ``parity kicks''. We also show 
that vibrational decoherence can be suppressed by this decoupling 
procedure, even though this is generally more difficult because the 
rate at which the parity kicks have to applied increases with the 
effective bath temperature. 
\end{abstract}


\section{Introduction}

Real world quantum systems interact with their environment to a greater 
or lesser extent. No matter how weak the coupling with such an 
environment, the evolution of an open quantum system is eventually 
affected by nonunitary features like decoherence, dissipation, and 
heating. Decoherence, in particular, is a 
serious obstacle to all applications exploiting quantum coherence, 
such as the bourgeoning field of quantum information processing.

Recently, considerable effort has been devoted to designing strategies 
able to counteract the undesired effects of the coupling with an
external environment. Notable examples of these strategies in the 
field of quantum information are quantum error correction codes
\cite{qecc} and error avoiding codes \cite{eac}, both based on 
encoding the state to be protected into 
carefully selected subspaces of the joint Hilbert space of the system 
and a number of ancillary systems. The main difference 
between the two encoding strategies is that error avoiding codes 
(also called decoherence-free subspaces) provide a passive strategy 
relying on the occurrence of specific symmetries in the interaction 
with the environment, which guarantees the existence of state space 
regions inaccessible to noise. Quantum error correction is instead an 
active strategy in which the encoding is performed in such a way that 
the various errors are mapped onto orthogonal subspaces so that they 
can be diagnosed and reversed.

A simple example of decoherence-free subspace has been recently 
demonstrated with two trapped ions \cite{kielp}, while error correction 
codes for single qubit errors 
has been demonstrated only in NMR quantum information processors 
\cite{nmrqecc}.
The main limitation for the efficient implementation of these encoding strategies
for combatting decoherence is the 
large amount of extra space resources required \cite{steane}. Correcting all 
the possible one-qubit errors requires at least five qubits 
\cite{zurprl} and if 
fault tolerant error correction is also considered, the number of 
ancillary qubits rapidly increases. For this reason, other alternative 
approaches which do not require any ancillary resources have been 
pursued, and which may be divided into two main categories: 
closed-loop (quantum feedback) \cite{closed,jmr},
and open-loop \cite{viola1,ban,noi,viola2,agarwal,berman} decoherence control 
strategies. In closed loop techniques, the system
to be protected is subject to appropriate measurements and the 
classical information obtained from this measurement is used for 
real-time correction of the system dynamics. This technique shares 
therefore some similarities with quantum error correction, which also 
checks which error has taken place and eventually corrects it. 
However, the main limiting aspect of feedback schemes is the need of 
a measurement, which is always inevitably subject to the limitations
due to non-unit detection efficiency. In fact, only under specific cases 
(see \cite{jmr}) it is possible
to automatically correct the error without a measurement, as in 
quantum error correction codes. In open loop control strategies
instead, the system is subject to external, suitably tailored, 
time-dependent drivings which are independent of the system dynamics 
and do not require any measurement, but only a limited, {\em a priori}, 
knowledge of the system-environment dynamics. These external control 
Hamiltonians are chosen in order to realize an effective 
dynamical decoupling of the system from the environment. In this way, 
any undesired effect of the environment, such as dissipation, 
decoherence, heating, can be eliminated in principle. The essential 
physical idea behind these open loop schemes 
comes from refocusing techniques in NMR spectroscopy,
now routinely used to eliminate unwanted interactions \cite{nmr}. Nonetheless,
these decoupling methods have recently attracted a large interest and 
they have been applied in many different situations, such as the 
inhibition of the decay of an unstable atomic state \cite{agarwal}, or
the suppression of magnetic state decoherence \cite{berman}.
The general applicability of decoupling methods has been discussed in 
\cite{viola2}, while the possibility to combine decoupling 
techniques together with weak-strength and slow-switching controls has been 
analysed in \cite{viola3}, where the conditions under which 
noise-tolerant, universal quantum control of a system can be performed 
with no extra space resources, have been determined. The general 
algebraic structure behind decoupling strategies has been also 
analysed in Ref.~\cite{zana1}, where it is shown how decoupling 
can be also considered as a dynamical {\em symmetrization} with 
respect to a group. This more general algebraic framework has also 
provided a unifying picture for coding and decoupling 
noise control strategies \cite{zana2,viola4}. In fact, when decoupling 
open loop controls are combined together with encoding into larger 
Hilbert spaces, fault-tolerant universal control of quantum systems 
becomes possible even with limited control resources. For example, it has been 
shown that the Heisenberg exchange interaction is 
sufficient to perform universal quantum computation if appropriately 
encoded qubits are used \cite{lidar}; these encoded subspaces may 
actually be made decoherence-free if appropriate decoupling controls are 
applied in parallel \cite{viola4,lidar2}.

The main drawback of open loop decoupling procedures is that the 
timing constraints are particularly stringent. In fact, the decoupling 
interactions has to be turned on and off at extremely short time 
scales, even faster than typical environmental timescale 
({\em full-strength/fast-switching} or {\em quantum bang-bang} controls
\cite{viola2}). In fact, perfect decoupling from the environment is 
obtained only in the infinitely fast control limit (see Section II)
and it is therefore important to establish in a quantitative way how 
effective these decoupling schemes are in a realistic 
situation with control pulses with finite strength and time duration.
A detailed analysis of decoupling timescales
has been performed only in \cite{viola1} 
for the case a single qubit in the presence of a purely dephasing 
environment, and in \cite{noi} in the case of a linearly damped
vibrational degrees of freedom. In this latter case, Ref.~\cite{noi}
proved that perfect decoupling can be achieved
using extremely fast ``parity kicks'', and that significant 
suppression of dissipation and decoherence due to the coupling with a 
zero-temperature bath is obtained as soon as the frequency of 
parity kicks becomes larger than the frequency cutoff of the 
environment. In the present paper we shall reconsider the model 
of Ref.~\cite{noi} and extend the analysis to the case of a {\em finite
temperature} environment. The motivation for this study is twofold. 
First of all it will allow us to establish if and how thermal effects 
influence the decoupling strategy, that is, if temperature introduces a new 
timescale which, together with the environmental frequency cutoff, 
determines the effectiveness of the parity kick decoupling strategy.
Secondly, the damped harmonic oscillator in a finite temperature bath 
studied in this paper well describes a collective vibrational
mode of a chain of trapped ions, which is used as a quantum bus
in linear ion trap quantum 
computers \cite{cizo}. One of the main experimental problems for 
quantum information processing with linear ion traps is just  
heating of these vibrational modes \cite{nist}, and it is therefore 
extremely important to establish if the parity kick decoupling method of 
Ref.~\cite{noi} is able to suppress heating and 
decoherence in this case.

The paper is organized as follows: In Section II decoupling strategies
in general,
and the parity kick method of Ref.~\cite{noi} as a particular example, 
are presented. 
In Section III the dynamics of the vibrational mode in the presence 
of a nonzero temperature bath and parity kicks is analyzed in detail 
and in Section IV the numerical results for 
both heating and decoherence rates are presented. 
Section V is for concluding remarks.

\section{Dynamical decoupling via parity kicks}

The starting point of decoupling techniques is the observation that 
even though one does not have access to the large number of 
uncontrollable degrees of freedom of the environment, it is still 
possible to interfere with its dynamics by inducing motions into 
the {\em system}, which are at least as fast as the environment 
dynamics. This indirect influence of the environment is obtained 
through the application of suitable time-dependent perturbations 
acting on the system variables only. Let us now review the main points 
of the decoupling technique following the lines of Ref.~\cite{viola2}.

We consider a quantum system $S$ coupled to an 
arbitrary bath $B$, whose overall Hamiltonian can be written as
\begin{equation}
H_0=H_S \otimes \openone_B + \openone_S \otimes H_B + H_{SB}=
\sum_{\alpha} {\cal S}_\alpha \otimes {\cal B}_\alpha \;. 
\label{hamiltonian}
\end{equation}
A decoupling strategy consists in trying to protect the evolution of 
$S$ against the effect of the
interaction $H_{SB}$, by seeking a perturbation $H_1(t)\otimes \openone_{B}$
to be added to $H_0$ so that the total Hamiltonian becomes 
$H(t)= H_0 + H_1(t) \otimes \openone_{B}$.
One usually restricts to situations where the
control field is {\em cyclic}, {\it i.e.}, associated to a decoupling
operator $U_1(t)$ that is periodic over some cycle time $T_c$:
\begin{equation}
U_1(t)\equiv T\exp\bigg\{ \hspace{-0.5mm} -(i/\hbar) \int_0^t\, du \,H_1(u)
\hspace{-0.5mm} \bigg\} = U_1(t+T_c) \;, 
\label{cyclicity}
\end{equation}
where $T$ denotes time ordering.
In this case one focus on the {\em stroboscopic} evolution at times 
$T_N=N T_c$, and it is possible to see that in this case the evolution
is driven by an effective {\em average} Hamiltonian \cite{waugh}
\begin{equation} 
U_{tot}(T_N) = \mbox{e}^{-(i/\hbar) \overline{H} T_N} 
\;. \label{propagator}
\end{equation}
The calculation of the average Hamiltonian
$\overline{H}$ is performed on the basis of a standard Magnus expansion of the
time-ordered exponential defining the cycle propagator \cite{wilcox}, 
\begin{equation}
U_{tot}(T_c) 
=\exp{ ( -i \overline{H} T_c/\hbar) }=
T\exp \bigg\{ \hspace{-0.5mm} -(i/\hbar) \int_0^{T_c} du 
\, \tilde{H}(u)  \hspace{-0.5mm} \bigg\} =
\mbox{e}^{-i[\, \overline{H}^{(0)} + \overline{H}^{(1)} + \ldots\,]\, 
T_c/\hbar} \;, 
\label{magnus}
\end{equation}
where 
\begin{equation}  
\tilde{H}(t) = U_1^\dagger (t) H_0 U_1(t) =
\sum_\alpha \Big[ U_1^\dagger (t) {\cal S}_\alpha U_1(t) \Big] \otimes 
{\cal B}_\alpha \;.    
\label{htilde}
\end{equation}
The various contributions in the right hand side of Eq.~(\ref{magnus})
collect terms of equal order in $\tilde{H}(t)$. In particular, 
\begin{eqnarray} 
\overline{H}^{(0)} & = & {1 \over T_c} \int_0^{T_c} du \, \tilde{H}(u)
\;, \label{zeroth} \\
\overline{H}^{(1)} & = & - {i \over 2 T_c} \int_0^{T_c} dv 
\int_0^{v} du \, \Big[ \tilde{H}(v), \tilde{H}(u) \Big] \;.
\label{first} 
\end{eqnarray}
One says that $k$th-order decoupling is achieved if the control field 
$H_1(t)$ can be devised so that contributions mixing $S$ and $B$ degrees of
freedom are no longer present in $\overline{H}^{(0)}$ and the first
nonvanishing correction arises from $\overline{H}^{(k)}$, $k \geq 1$. 
One then considers the infinitely fast control limit,
which, for a finite evolution time $T$, requires considering
$T_c=T/N$ in the limit $T_{c} \rightarrow 0$ and $N\rightarrow\infty$.
In this limit, first-order decoupling is sufficient,
contributions higher 
than zeroth-order are negligible in (\ref{magnus}), and one can focus on the 
problem of designing the effective Hamiltonian $\overline{H}^{(0)}$
of Eq.~(\ref{zeroth}) in such a way that there is no residual 
system-environment coupling. 

The more general way to engineer the average Hamiltonian $\overline{H}^{(0)}$
is through {\em symmetrization} with respect to a finite group ${\cal 
G}$ \cite{viola2,zana1}. In fact, if we consider a finite group of 
unitary operators ${\cal G} = \{g_{j}\}, j=1,\ldots , |{\cal G}|$, 
symmetrization is the map (acting on system operator only)
\begin{equation}
{\cal S} \mapsto \Pi_{\cal C}({\cal S}) =
 {1 \over |{\cal G}| } 
\sum_{g_j \in {\cal G}} \, g_j^\dagger\, {\cal S} \, g_j,
\label{map}
\end{equation}
which is also the projection on the so-called {\sl centralizer} of 
${\cal G}$, composed of operators 
commuting with every element $g_j$ of the group ${\cal G}$ \cite{viola2,zana1}.
The map (\ref{map}) 
can be dynamically implemented through a simple 
piecewise constant decoupling operator:
\begin{equation}
U_1(t) \equiv g_j\;, \hspace{5mm} j \,\Delta t \leq t < (j+1) \,\Delta t \;, 
\label{decoupling}
\end{equation}
corresponding to a partition of the cycle time $T_c$ into $|{\cal G}|$ 
intervals of equal length $\Delta t \equiv T_c /|{\cal G}|$. Then, by 
(\ref{htilde}), 
\begin{equation}
\overline{H}^{(0)} = \Pi_{\cal C} (H_0) = \sum_\alpha \, 
\Pi_{\cal C}({\cal S}_\alpha) \otimes {\cal B}_\alpha \;, 
\label{proofa}
\end{equation}
showing that the average Hamiltonian $\overline{H}^{(0)}$, generating 
time evolution in the infinitely fast control limit, has been 
symmetrized, i.e., has become invariant with respect to the group ${\cal G}$.
Perfect decoupling from the environment is achieved when 
\begin{equation}
\Pi_{\cal C} (H_{SB})=0\;,
\label{decoup}
\end{equation}
and in this case the effective open system evolution for the reduced 
density operator of the system $\rho_{S}$ over time $T$ is governed by
\begin{equation}
\lim_{N \rightarrow \infty} \rho_S(T=NT_c) = e^{-i \overline{H}_S T/\hbar }
\,\rho_S(0)\, e^{+i \overline{H}_S T/\hbar } \;, 
\label{select}  
\end{equation} 
where $\overline{H}_S=\Pi_{\cal C}(H_S)$. This means that the system 
is no more interacting with the environment and the residual time 
evolution is driven by a projected system Hamiltonian,
invariant with respect to ${\cal G}$.

A number of examples of decoupling groups ${\cal G}$ has now 
appeared in the literature, expecially for the case of many-qubits 
dissipative registers with various kinds of interaction with the 
environment \cite{viola1,ban,viola2,zana1}. Another important example 
for applications in quantum computing is the case of a 
linearly dissipative vibrational degree of freedom, which can be used 
as a {\em quantum bus} in linear ion trap quantum computers \cite{cizo}, 
and which has been 
shown in Ref.~\cite{noi} to be decoupled by the group ${\cal Z}_{2}$, 
composed by the identity and the parity operator $P$. In fact, it is 
straightforward to check that the decoupling condition (\ref{decoup})
is equivalent to the condition of Eq.~(7) of Ref.~\cite{noi}. 

As it can be expected, implementing the above general decoupling strategy 
is by no means trivial. First of all, for a given $H_{SB}$, the 
identification of a minimal group ${\cal G}$ able to produce 
decoupling is nontrivial. Secondly,
the decoupling prescription (\ref{decoupling}) requires the 
capability of instantaneously changing the evolution operator from $g_j$ 
to $g_{j+1}$ over successive subintervals. This means assuming the 
capability of implementing arbitrarily strong and extremely fast 
control operations.
Such impulsive full-power control configurations correspond to 
so-called {\em quantum bang-bang controls} as introduced in \cite{viola1}. 
As it has been already shown in \cite{viola1,noi}, the most stringent 
condition is not on the strength but rather on 
the extremely high speed of the control operations: one 
has to be faster than the typical {\em environmental} timescale, 
which is usually fixed by the frequency cutoff of the bath  
spectrum, $\omega_{c}$. However, the identification of the frequency 
cutoff $\omega_{c}$ as the only relevant parameter determining the 
threshold for the decoupling cycle frequency $1/T_{c}$ above which the 
decoupling procedure becomes effective, has been done in 
Refs.~\cite{viola1,noi} only on the basis of two specific examples. 
In Ref.~\cite{noi}, the case of a harmonic oscillator coupled to a 
zero-temperature bath, able to induce only system dissipation (and the 
associated decoherence) has been considered. 
The case of nonzero temperature has been discussed in 
Ref.~\cite{viola1} but only in the particular case of a single qubit 
subject to a purely dephasing, energy-conserving, environment. It is 
therefore important to establish the effectiveness of decoupling 
techniques in the general case of a {\em nonzero-temperature, 
dissipative bath}. From now on we shall specialize to the case of the 
linearly damped harmonic oscillator of Ref.~\cite{noi}, which is of 
relevance for linear ion trap quantum computation. In fact, we 
shall demonstrate that decoupling techniques can be successfully used
to efficiently suppress heating of the vibrational center-of-mass motion
of the ion chain.

\section{Parity kicks for a damped harmonic oscillator} 

We choose a harmonic oscillator as system of interest 
\begin{equation}
H_{S}=\hbar \omega _{0} a^{\dagger} a  \;,
\label{ha}
\end{equation}
describing a collective vibrational mode of a linear chain
of trapped ions with frequency $\omega_{0}$.
It has been already experimentally verified \cite{nist,nat}
that the nonunitary features of the vibrational dynamics (heating and 
decoherence) are well described by modelling the
environment as a collection of independent bosonic modes 
\cite{caldleg} 
\begin{equation}
H_{B}=\sum_{k}\hbar \omega _{k} b_{k}^{\dagger} b_{k} \;, 
\label{hb}
\end{equation}
interacting with the vibrational mode via the following bilinear 
interaction Hamiltonian in which the ``counter-rotating'' terms are dropped
\begin{equation}
H_{SB}=\sum_{k} \hbar \gamma_{k} \left(ab_{k}^{\dagger}
+a^{\dagger}b_{k}\right)  \;.
\label{hint2}
\end{equation}
The symmetrization with respect to the group ${\cal Z}_{2}$ is 
performed by periodically pulsing the oscillation frequency, that is, 
by changing the potential so that $\omega_{0}$ is changed to $\omega_{0}+\delta 
\omega $ for a time interval $\tau$ and with a time period $T_{c}$ 
(see Fig.~1). The pulse realizes the ``parity kick'' of Ref.~\cite{noi}
when the condition $\delta \omega \cdot \tau = \pi$ is 
satisfied.
In this way, the cyclic time-dependent control Hamiltonian is given by
\begin{equation}
H_{1}(t)= \hbar \delta \omega a^{\dagger} a \sum_{n=1}^{\infty}
\theta\left(t- nT_{c}+\tau)\right)
\theta\left(nT_{c}-t \right)\;,
\label{ki2}
\end{equation}
so that the cyclic decoupling operator $U_{1}(t)$ is equal to 
$U_{1}(t)= \openone_{S}$ for $0<t<T_{c}-\tau$ and 
$U_{1}(t)= \exp\{i\pi a^{\dagger} a\} = P$, for $T_{c}-\tau < 
t < T_{c}$.

\begin{figure}[bt]
\centerline{\epsfig{figure=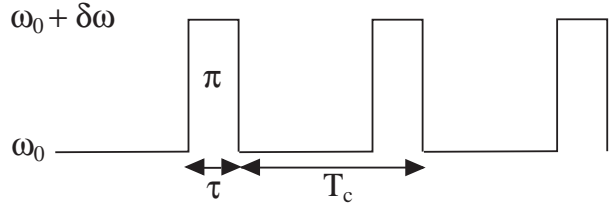,height=4cm}}
\caption{Sketch of the implementation of the parity kick decoupling 
procedure by pulsing the oscillation frequency.}
\label{fig1}
\end{figure}

To determine the effects of a nonzero temperature bath 
on the efficiency of the decoupling scheme, we shall 
consider an initially factorized state in which the vibrational mode 
is prepared in a given pure state $|\psi(0)\rangle $ and 
the environment is at the thermal equilibrium state at temperature 
$T$, $\rho_{B}^{T}$. To be more specific we want to establish if 
decoupling via parity kicks is able to suppress efficiently both  
heating (which is important for quantum 
information processing) and quantum decoherence of the vibrational mode.
To study heating we shall assume that the collective 
vibrational mode has been initially cooled to its ground state \cite{groundcool}, 
that is, $|\psi(0)\rangle = |0\rangle $. The study of 
decoherence instead will be performed, as in 
Ref.~\cite{noi}, by considering an 
initial linear superposition of two coherent states with opposite 
phases, that is, the 
well known Schr\"odinger cat state
\begin{equation}
|\psi(0)\rangle=
|\psi_{\varphi}\rangle = N_{\varphi}\left(|\alpha(0)\rangle +
e^{i\varphi}|-\alpha(0)\rangle\right) \;,
\label{cat}
\end{equation}
where 
$N_{\varphi}=(2+2e^{-2 |\alpha(0)|^{2}}\cos\varphi)^{-1/2}$.
The dynamics of the system in the presence of parity kicks
will be exactly solved for both initial conditions of 
the vibrational mode, by exploiting the fact that 
a tensor product of coherent states retains its 
form at all times when the evolution is generated by the Hamiltonian
of Eqs.~(\ref{ha}), (\ref{hb}) and (\ref{hint2}) \cite{romero}, that is
\begin{equation}
	|\alpha(0)\rangle \otimes \prod_{k}|\beta_{k}(0)\rangle 
	\rightarrow
	|\alpha(t)\rangle \otimes \prod_{k}|\beta_{k}(t)\rangle \;,
	\label{evocohe}
\end{equation}
where the time-dependent coherent state amplitudes are a linear combination of 
the initial amplitudes
\begin{eqnarray}
	\alpha (t) & =& L_{00}(t)\alpha(0)+\sum_{k}L_{0k}(t)\beta_{k}(0) 
	\label{eql1} \\
	\beta_{k} (t) & =& L_{k0}(t)\alpha(0)+\sum_{k'}L_{kk'}(t)\beta_{k'}(0).  
	\label{eql2}
\end{eqnarray}
Eq.~(\ref{evocohe}) is useful also in the nonzero temperature
case. In fact, using the expression 
of the thermal state $\rho_{B}^{T}$ in the Glauber-Sudarshan 
P-representation \cite{qnoise}
\begin{equation}
\rho_{B}^{T}=\prod_{k}\int \frac{d^{2}\beta_{k}}{\pi 
N_{k}}\exp\left\{-\frac{|\beta_{k}|^{2}}{N_{k}}\right\}|\beta_{k}\rangle 
\langle \beta_{k}|,
\label{pther}
\end{equation}
where $d^2\beta_k =d{\rm Re}\beta_k d{\rm Im}\beta_k$ and
$N_{k}=(\exp\left\{\hbar \omega_{k}/k_{B}T\right\}-1)^{-1}$
is the mean thermal excitation number of the k-th bath mode,
one has always to evaluate the time evolution of terms like 
$|\alpha\rangle
\langle \alpha '||\beta_{k}\rangle \langle \beta_{k}|$, and then 
perform the average over the thermal Gaussian weight 
$\exp\left\{-|\beta_{k}|^{2}/N_{k}\right\}/\pi N_{k}$.
Therefore the essential dynamics is contained in the expression of the 
unitary matrix $L_{ij}(t)$ of Eqs.~(\ref{eql1}) and (\ref{eql2}),
which has the same 
structure both with and without parity kicks, because the two 
situations differ only by the value of the oscillation frequency.
The matrix element $L_{00}(t)$ is given in terms of its Laplace transform, 
and, in the interaction picture with respect to $H_{S}$ of Eq.~(\ref{ha}),
one has 
\begin{equation}
L_{00}(t,\delta \omega)=\mathcal{L}\mathit{^{-1}\left[\frac{1}
{z+K(z,\delta \omega)}\right
]} \;,
\label{azz} 
\end{equation}
where 
\begin{equation}
K(z,\delta \omega)= \sum_{k} \frac{\gamma_{k}^{2}}{z+i(\omega_{k}-\omega_{0})
-i\delta \omega} \;.
\label{kapa}
\end{equation}
This expression refers to the evolution during the parity kicks, that 
is, for $nT_{c}-\tau < t < nT_{c}$, $n\geq 1$. The evolution in the absence of 
kicks is simply obtained putting $\delta \omega 
=0$ in Eqs.~(\ref{azz}) and (\ref{kapa}). 
All the other matrix elements can be 
expressed in terms of the matrix element $L_{00}(t,\delta \omega)$
in the following way:
\begin{equation}
L_{0k}(t,\delta \omega)=L_{k0}(t,\delta \omega) 
=-i \gamma_{k}\int_{0}^{t}ds e^{-i(\omega_{k}-\omega_{0}-
\delta \omega )s}
L_{00}(t-s,\delta \omega)  
\label{azk}
\end{equation}
\begin{equation}
L_{kk'}(t,\delta \omega)=\delta_{kk'}e^{-i(\omega_{k}-\omega_{0})t} 
 - \gamma_{k}\gamma_{k}'\int_{0}^{t}ds 
e^{-i(\omega_{k}-\omega_{0}-\delta \omega)(t-s)}\int_{0}^{s}ds' 
e^{-i(\omega_{k'}-\omega_{0}-\delta \omega)(s-s')}
L_{00}(s',\delta \omega)\;.   
\label{akk}
\end{equation}

It is evident that a decoupling cycle of duration $T_{c}$ will be 
described by the product of unitary matrices $L(\tau,\delta \omega)
\cdot L(T_{c}-\tau,0)$ applied to the vector formed by the 
coherent amplitudes $(\alpha(t), \ldots\ldots \beta_{k}(t)\ldots..)$. 
As a consequence, the stroboscopic dynamics of the whole system
during the decoupling procedure, in 
the case of an initial tensor product of coherent states, can be 
described exactly as
\begin{equation}
	\left(\matrix{
		\alpha(NT_{c}) \cr
		\vdots \cr
		\beta_{k}(NT_{c})  \cr
		\vdots \cr}\right)  
		= \left[L(\tau,\delta \omega)\cdot L(T_{c}-\tau,0) 
\right]^{N}
		\left(\matrix{
		\alpha(0) \cr
		\vdots \cr
		\beta_{k}(0)  \cr
		\vdots \cr}\right)  \; .
\label{matkik}
\end{equation}

\section{Numerical results}

In the standard description of dissipation, one always considers a 
continuum distribution of oscillator frequencies in order to obtain 
an irreversible transfer of energy from the system of interest into the
reservoir. Moreover, most often, also the Markovian assumption is made
which means assuming an infinitely fast bath with an infinite 
frequency cutoff $\omega_{c}$. This case of a standard vacuum bath
in the Markovian limit is characterized by an infinite, continuous 
and flat distribution of couplings \cite{qnoise},
\begin{equation}
	\gamma(\omega)^{2}=\frac{\gamma}{2 \pi} \;\;\;\;\forall \omega \;,
	\label{fla}
\end{equation}
where $\gamma$ is the energy damping rate.
As shown in Refs.~\cite{viola1,noi}, decoupling strategies become 
efficient when the external controls are characterized by timescales 
faster than those of the environment. It is therefore evident that, in 
the presence of parity kicks, we cannot make any Markovian 
approximation. We have to solve
numerically the problem, by simulating the continuous distribution
of bath oscillators with a large but finite number of oscillators
with closely spaced frequencies. As in Ref.~\cite{noi}, we have 
considered a bath of $201$ oscillators, with equally spaced 
frequencies, symmetrically distributed around the resonance frequency
$\omega_{0}$, i.e.
\begin{eqnarray}
&&\omega_{k}=\omega_{0}+k \Delta \;\;\;\;\;\;\;\; 
\Delta=\frac{\omega_{0}}{100} \label{omk}\\
&& k_{max}=\frac{\omega_{0}}{\Delta}=100 \Rightarrow \omega_{k}^{max}=
2 \omega_{0} \\
&& k_{min}=-k_{max}=-100 \Rightarrow \omega_{k}^{min}=0 \;, \\
\end{eqnarray}
and we have considered a constant distribution of couplings
similar to that associated with the Markovian limit
\begin{equation}
\gamma_{k}^{2}=\frac{\gamma \Delta}{2\pi} \;\;\;\;\;\;\forall k \;.
\label{gak}
\end{equation}
Approximating a continuous Markovian bath
with a finite number of bath oscillators has two main effects.
First of all, the discrete frequency distribution with a fixed 
spacing $\Delta$ makes all the dynamical quantities
periodic with period $T_{rev}=2\pi/\Delta$ \cite{cukier}. Therefore
our numerical solution will correctly
describe the interaction with the environment provided that we consider 
not too large times, say $t \leq \pi/\Delta $. 
Secondly, the introduction of a finite cutoff ($\omega_{c}=
2\omega_{0}$ in our case) implies a modification of the 
coupling spectrum $\gamma(\omega)$ at very high frequency with respect
to the infinitely flat distribution of the Markovian treatment (see
Eq.~(\ref{fla})).
This fact manifests itself in a slight modification of the
dynamics at very short times ($t \simeq \omega_{c}^{-1}$) \cite{cukier}.
We have verified both short and long time deviations from 
the standard Markovian bath dynamics in our numerical calculations.
However, we have checked that our model environment with a finite
number of oscillators faithfully reproduces the standard Markovian 
bath dynamics within the
time interval of interest, $0.1 /\gamma < t < 3/\gamma $ say.

\subsection{Effect of parity kicks on heating}

To check if decoupling via 
parity kicks is able to suppress the heating of the vibrational mode, 
we consider the 
following initial state for the whole system
\begin{equation}
|0\rangle \langle 0| \otimes \prod_{k}\int \frac{d^{2}\beta_{k}(0)}{\pi 
N_{k}}\exp\left\{-\frac{|\beta_{k}(0)|^{2}}{N_{k}}\right\}|\beta_{k}(0)\rangle 
\langle \beta_{k}(0)|,
\label{iniheat}
\end{equation}
where $|0\rangle $ is the ground state of the collective vibrational 
mode \cite{groundcool}.

Using the property (\ref{evocohe}), and tracing over the 
environment, the evolved state of the vibrational mode after $N$
decoupling cycles can be written as
\begin{equation}
\rho_{S}(NT_{c})=\int \prod_{k}\frac{d^{2}\beta_{k}(0)}{\pi 
N_{k}}\exp\left\{-\frac{|\beta_{k}(0)|^{2}}{N_{k}}\right\}
|\alpha(NT_{c})\rangle \langle \alpha(NT_{c})|,
\label{evoheat}
\end{equation}
where the coherent state amplitude $\alpha(NT_{c})$ 
is the following linear combination of the complex variables $\beta_{k}(0)$, 
\begin{equation}
\alpha(NT_{c}) = \sum_{k} C_{0k}(NT_{c})\beta_{k}(0),
\label{evoheat2}
\end{equation}
where we have defined $C_{0k}(NT_{c})=
\left\{\left[L(\tau,\delta \omega)\cdot L(T_{c}-\tau,0) 
\right]^{N}\right\}_{0k}$ 
(see Eqs.~(\ref{eql1}) and (\ref{matkik})). The Gaussian
average of Eq.~(\ref{evoheat}) can be performed by first considering the 
normally ordered characteristic function $\chi(\lambda, NT_{c})$ \cite{qnoise} 
of the state, and then performing the integration. One gets
\begin{equation}
\chi(\lambda, NT_{c})=\exp\left\{-|\lambda |^{2}\sum_{k} 
N_{k} |C_{0k}(NT_{c})|^{2}\right\},
\label{chara}
\end{equation}
showing that, in the presence of parity kicks,
the vibrational state is a {\em thermal 
state, with mean vibrational number} $\nu(NT_{c})$,
\begin{equation}
\nu (NT_{c})=\sum_{k} 
N_{k} |C_{0k}(NT_{c})|^{2}.
\label{meannum}
\end{equation} 
The stroboscopic time evolution of this mean vibrational number is plotted in 
Fig.~2 both in the presence (full circles) and in the absence 
(crosses) of parity kicks. The capability of the parity kick 
decoupling strategy to avoid vibrational heating is clearly visible
in this figure. In Fig.~2 and in the rest of the paper we consider a 
vibrational mode with frequency $\omega_{0}=10$ Mhz, damping rate 
$\gamma = 0.1 $ Mhz and environmental frequency cutoff $\omega_{c}=20$ 
Mhz. The curve referring to the situation without parity kicks in 
Fig.~2 well reproduces the standard Markovian result \cite{qnoise}
$\nu(t)=N(\omega_{0})(1-e^{-\gamma t})$, where 
$N(\omega_{0})=(\exp\left\{\hbar \omega_{0}/k_{B}T\right\}-1)^{-1}$ 
is the mean vibrational number of the oscillator at thermal 
equilibrium in the usual Born-Markov approximation. 
Fig.~2 refers to an effective reservoir temperature 
$T=10$ mK (corresponding to $N(\omega_{0}) \simeq 130$), and to the 
following decoupling cycle parameters: $T_{c}=157$ ns, parity kick 
duration $\tau=T_{c}/7\simeq 22.4$ ns, implying $\delta \omega =140$ Mhz.

\begin{figure}[bt]
\centerline{\epsfig{figure=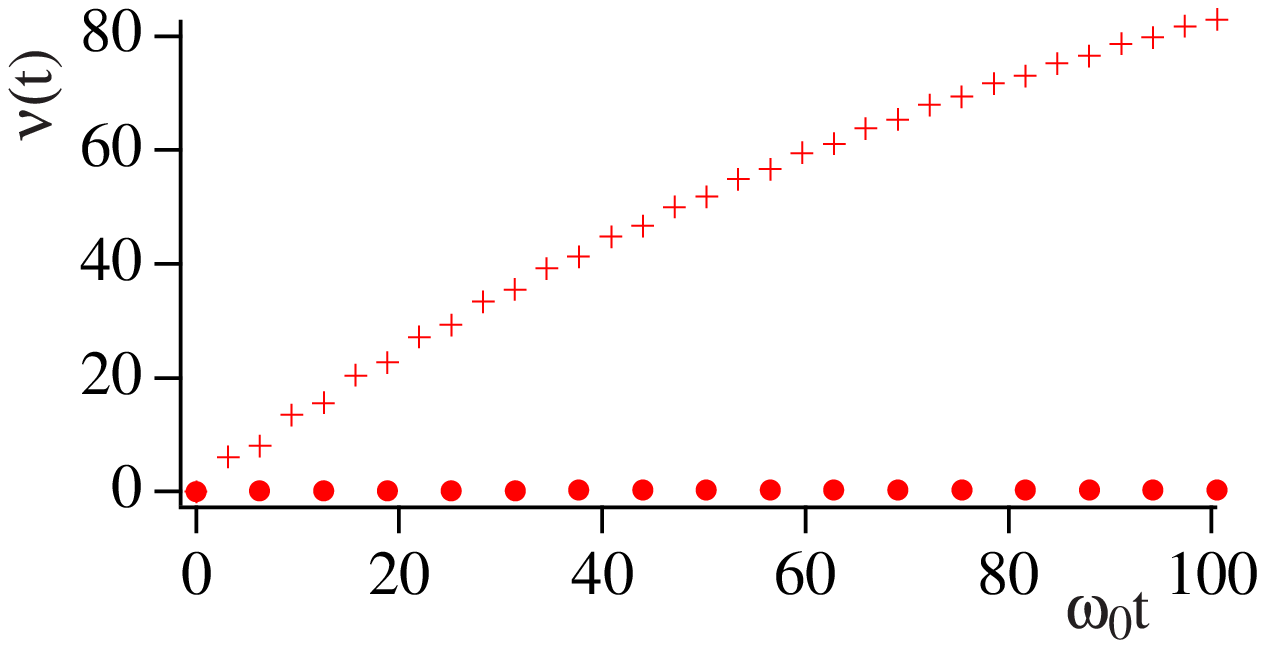,height=6cm}}
\caption{Time evolution of the mean vibrational number of 
Eq.~(\protect\ref{meannum}) with (full circles) and without (crosses)
parity kicks. The capability of parity kicks to suppress heating is 
clearly visible. Parameters are: $\omega_{0}=10$ Mhz, 
$\gamma = 0.1 $ Mhz, $\omega_{c}=20$ 
Mhz, effective reservoir temperature 
$T=10$ mK (corresponding to $N(\omega_{0}) \simeq 130$),
$T_{c}=157$ ns, parity kick 
duration $\tau=T_{c}/7\simeq 22.4$ ns, implying $\delta \omega =140$ 
Mhz.}
\label{fig2}
\end{figure}

The influence of the environmental temperature $T$ on heating suppression
is analysed in Fig.~3, where the mean vibrational 
number after one relaxation time $t=1/\gamma$, $\nu(1/\gamma)$, is 
plotted as a function of the rescaled decoupling cycle time 
$\omega_{c}T_{c}/2\pi$ for three different bath temperatures, $T=10$ 
mK (a), $T=100$ mK (b), and $T=1$ K (c). For each value of $T_{c}$, we 
have always chosen the kick duration $\tau =T_{c}/7$, as in Fig.~2,
and the frequency shift $\delta \omega $ is always correspondingly
adjusted so that $\delta \omega = \pi/\tau$. We can see that a well 
visible threshold for the decoupling cycle time $T_{c}$ exists and 
that as soon as the parity kicks are sufficiently fast,
$T_{c} < 2\pi/\omega_{c}$, heating suppression becomes significant.
This is a phase transition-like behavior analogous to that
found for decoherence suppression in the zero-temperature case \cite{noi}.
What is more important is that {\em bath temperature has no 
effect} on the effectiveness of the decoupling scheme: the results are 
essentially identical for the three different temperatures studied,
and this means that, at least for what concerns heating, the only relevant 
environmental timescale is given by the frequency cutoff $\omega_{c}$.
This result is particularly 
important for the application of the parity kick strategy to suppress 
heating in linear ion trap quantum computers, where heating is due to 
some technical imperfections originating from fluctuating patch fields 
\cite{nist}. Determining the effective temperature $T$ of the thermal 
bath modelling these fluctuating fields is generally very difficult,
but our results shows that this is not 
relevant, and that parity kick decoupling 
is very promising for eliminating vibrational heating.

\begin{figure}[bt]
\centerline{\epsfig{figure=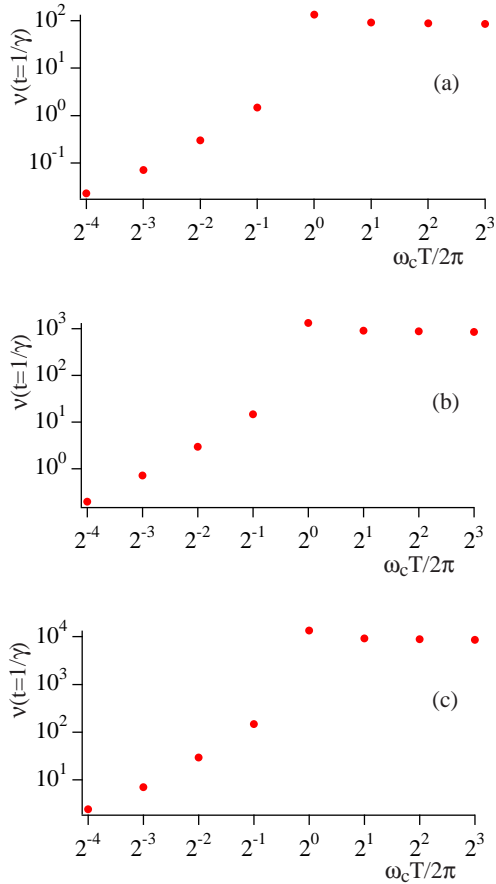,height=12cm}}
\caption{Mean vibrational 
number after one relaxation time $t=1/\gamma$, $\nu(1/\gamma)$,  
as a function of the rescaled decoupling cycle time 
$\omega_{c}T_{c}/2\pi$, for three different bath temperatures: $T=10$ 
mK (corresponding to $N(\omega_{0}) \simeq 130$) (a), $T=100$ mK 
(corresponding to $N(\omega_{0}) \simeq 1302$) (b), and $T=1$ K 
(corresponding to $N(\omega_{0}) \simeq 13144$) (c). 
For each value of $T_{c}$, we 
have always chosen $\tau =T_{c}/7$, and, correspondingly,
$\delta \omega = \pi/\tau$. The other parameters are as in Fig.~2}
\label{fig3}
\end{figure}

\subsection{Effect of parity kicks on decoherence}

Let us now consider the possibility to suppress decoherence.
We assume an 
initially prepared Schr\"odinger cat state of the  
vibrational mode, and therefore the following initial state for the whole
system
\begin{equation}
|\psi_{\varphi}\rangle \langle \psi_{\varphi}| 
\otimes \prod_{k}\int \frac{d^{2}\beta_{k}(0)}{\pi 
N_{k}}\exp\left\{-\frac{|\beta_{k}(0)|^{2}}{N_{k}}\right\}|\beta_{k}(0)\rangle 
\langle \beta_{k}(0)|,
\label{inidec}
\end{equation}
where $|\psi_{\varphi}\rangle $ is given by Eq.~(\ref{cat}). 

Using the property (\ref{evocohe}), and tracing over the 
environment, the evolved state of the vibrational mode after $N$
decoupling cycles can be written as
\begin{eqnarray}
&& \rho_{S}(NT_{c})=N_{\varphi}^{2}\int \prod_{k}\frac{d^{2}\beta_{k}(0)}{\pi 
N_{k}}\exp\left\{-\frac{|\beta_{k}(0)|^{2}}{N_{k}}\right\}\left\{
|\alpha_{+}(NT_{c})\rangle \langle \alpha_{+}(NT_{c})| +
|\alpha_{-}(NT_{c})\rangle \langle \alpha_{-}(NT_{c})|\right. 
\nonumber \\
&&\left. +e^{i\varphi}\langle \beta_{k}^{+}(NT_{c})|\beta_{k}^{-}(NT_{c})\rangle
|\alpha_{-}(NT_{c})\rangle \langle \alpha_{+}(NT_{c})| +
+e^{-i\varphi}\langle \beta_{k}^{-}(NT_{c})|\beta_{k}^{+}(NT_{c})\rangle
|\alpha_{+}(NT_{c})\rangle \langle \alpha_{-}(NT_{c})|\right\}.
\label{evodec}
\end{eqnarray}
The coherent state amplitudes $\alpha_{\pm}(NT_{c})$ and 
$\beta_{k}^{\pm}(NT_{c})$
are now given by the following linear combinations of the 
initial amplitudes (see Eqs.~(\ref{eql1})-(\ref{eql2}))
\begin{eqnarray}
&& \alpha_{\pm}(NT_{c}) = \pm \alpha_{0}C_{00}(NT_{c})+
\sum_{k} C_{0k}(NT_{c})\beta_{k}(0), \label{evodeclin1}\\
&& \beta_{k}^{\pm}(NT_{c}) = \pm \alpha_{0}C_{k0}(NT_{c})+
\sum_{k'} C_{kk'}(NT_{c})\beta_{k'}(0), 
\label{evodeclin2}
\end{eqnarray}
The Gaussian
average of Eq.~(\ref{evodec}) can be performed, as in the preceding subsection,
by first considering the 
normally ordered characteristic function 
of the state and then performing the integration. 
The integration is straightforward but lengthy, and the resulting 
reduced vibrational state may be better expressed in terms of its 
Wigner function $W_{S}(\alpha,NT_{c})$,
\begin{eqnarray}
&&W_{S}(\alpha,NT_{c})=\frac{2N_{\varphi}^{2}}{\pi\left[1+2\nu(NT_{c})\right]}
\left\{\exp\left\{-\frac{2|\alpha-\alpha_{0}C_{00}(NT_{c})|^{2}}
{1+2\nu(NT_{c})}\right\}+\exp\left\{-\frac{2|\alpha+\alpha_{0}C_{00}(NT_{c})|^{2}}
{1+2\nu(NT_{c})}\right\} \right. \nonumber \\
&& \left. +2\exp\left\{-2|\alpha_{0}^{2}|\eta(NT_{c})\right\}
\exp\left\{-\frac{2|\alpha|^{2}}
{1+2\nu(NT_{c})}\right\}\cos\left[\varphi +\frac{4 {\rm 
Im}\left[\alpha\alpha_{0}C_{00}(NT_{c})\right]}{1+2\nu(NT_{c})}\right]\right\},
\label{wign}
\end{eqnarray}
where $\nu(NT_{c})$ is again the mean vibrational number of the cat 
state of Eq.~(\ref{meannum}), the matrix element $C_{00}(NT_{c})$ 
describes the amplitude decay, and $\eta(NT_{c})$ is the 
{\em fringe visibility function} \cite{kenwal},
determining the relative strength of the
quantum interference term in the cat state, and which can be expressed 
as
\begin{equation}
\eta(NT_{c})= 1-\frac{|C_{00}(NT_{c})|^{2}}{1+2\nu(NT_{c})}.
\label{etastr}
\end{equation}
This fringe visibility is always contained in the interval $[0,1]$ and 
provides a good quantitative description of dynamical decoherence 
processes. For this reason we shall study the stroboscopic evolution 
of this quantity, as in Ref.~\cite{noi}, to quantify the eventual 
decoherence suppression caused by the decoupling.

The time evolution of the fringe visibility is plotted in 
Fig.~4, both with (full circles) and without 
(crosses) parity kicks. The possibility to suppress decoherence
using parity kicks is clearly demonstrated
in this figure. Parameters are the same as in Fig.~2, except that
the decoupling cycle parameters now are: $T_{c}=78.5$ ns, parity kick 
duration $\tau=T_{c}/7\simeq 11.2$ ns, implying $\delta \omega =280$ Mhz.
The curve referring to the situation without parity kicks in 
Fig.~4 (crosses) well reproduces the Markovian result 
\cite{kenwal}
\begin{equation}
\eta(t)=1-\frac{e^{-\gamma t}}{1+2N(\omega_{0})(1-e^{-\gamma t})}
\label{etaterm}
\end{equation}

\begin{figure}[bt]
\centerline{\epsfig{figure=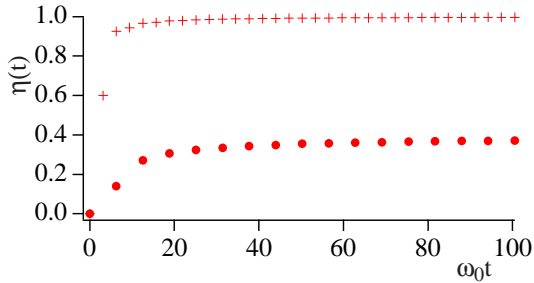,height=4cm}}
\caption{Time evolution of the fringe visibility function of 
Eq.~(\protect\ref{etastr}) with (full circles) and without (crosses)
parity kicks. The capability of parity kicks to suppress decoherence is 
clearly visible. Parameters are the same as in Fig.~2, except that
the decoupling cycle parameters now are: $T_{c}=78.5$ ns, parity kick 
duration $\tau=T_{c}/7\simeq 11.2$ ns, implying $\delta \omega =280$ Mhz.}
\label{fig4}
\end{figure}

The influence of temperature on decoherence suppression
is studied in Fig.~5, where the fringe visibility 
function after one relaxation time $t=1/\gamma$, $\eta(1/\gamma)$, is 
plotted as a function of the rescaled decoupling cycle time 
$\omega_{c}T_{c}/2\pi$ for three different bath temperatures, $T=10$ 
mK (a), $T=100$ mK (b), and $T=1$ K (c). For each value of $T_{c}$, we 
have always chosen the kick duration $\tau =T_{c}/7$, as in Figs.~2 
and 3,
and the frequency shift $\delta \omega $ is always correspondingly
adjusted so that $\delta \omega = \pi/\tau$. 

We can see from Fig.~5 that the situation is rather different 
from that with heating suppression. In fact, 
decoherence suppression by parity kicks {\em strongly depends on the 
bath temperature}, and it is significant only in the lower temperature 
case (Fig.~5a), which is the only case in which a threshold for the 
decoupling cycle time $T_{c}$ at about $T_{c} \simeq 2\pi/\omega_{c}$,
as in the zero temperature case \cite{noi}, is visible. In the other 
cases, decoherence suppression worsens for increasing bath temperature.
This result shows that eliminating decoherence via decoupling 
techniques is generally more difficult than eliminating heating. This 
can be easily explained in terms of the so-called 
thermal acceleration of decoherence \cite{kenwal,goetsch}, 
that is, the fact that in the case of a thermal bath at temperature $T$,
the decoherence process is accelerated roughly 
by a factor $(1+2N(\omega_{0}) )$ with respect to the zero 
temperature case. This thermal effect on the decoherence rate 
can be also easily checked from the Markovian 
limit expression of Eq.~(\ref{etaterm}). In fact, the fringe 
visibility function $\eta(t)$ reaches its asymptotic value in a time 
of the order of $t_{dec} \simeq [\gamma (1+2N(\omega_{0}) )]^{-1}$, 
and it is evident that decoherence suppression with parity kicks is 
possible only if the cycle time $T_{c}$ is smaller than this 
decoherence time $t_{dec}$, and not only smaller than 
$2\pi/\omega_{c}$, 
as in the zero temperature case. This means that, in a nonzero 
temperature bath, one has a new, {\em temperature-dependent}, 
threshold for decoherence suppression, given by
\begin{equation}
T_{c} < {\rm min}\left\{2\pi/\omega_{c},[\gamma (1+2N(\omega_{0}) 
)]^{-1}\right\},
\label{thres}
\end{equation}
and this generalized expression easily explains the results of Fig.~5. 

\begin{figure}[bt]
\centerline{\epsfig{figure=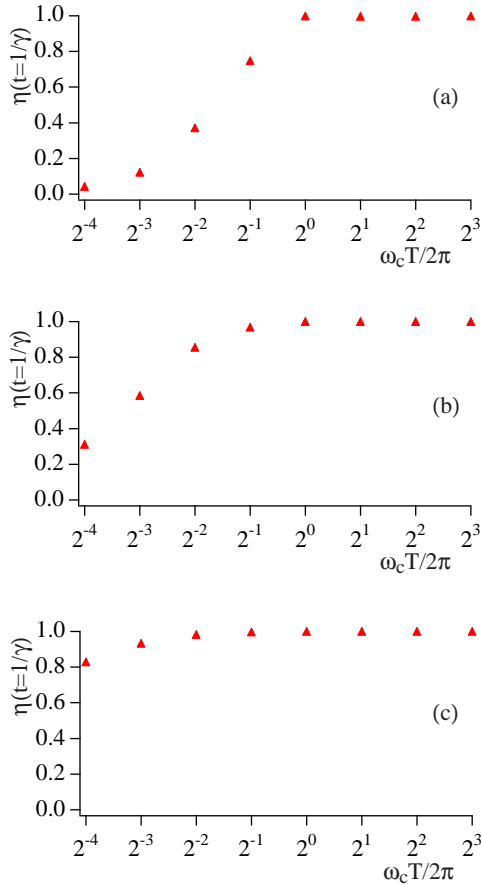,height=12cm}}
\caption{Fringe visibility function after one 
relaxation time $t=1/\gamma$, $\eta(1/\gamma)$,  
as a function of the rescaled decoupling cycle time 
$\omega_{c}T_{c}/2\pi$, for three different bath temperatures: $T=10$ 
mK (corresponding to $N(\omega_{0}) \simeq 130$) (a), $T=100$ mK 
(corresponding to $N(\omega_{0}) \simeq 1302$) (b), and $T=1$ K 
(corresponding to $N(\omega_{0}) \simeq 13144$) (c). 
For each value of $T_{c}$, we 
have always chosen $\tau =T_{c}/7$, and, correspondingly,
$\delta \omega = \pi/\tau$. The other parameters are as in Fig.~2.
The quality of decoherence suppression degrades with increasing 
temperature.}
\label{fig5}
\end{figure}

\section{Conclusions}

We have studied the application of open-loop decoupling schemes in an 
experimentally realistic scenario. In fact, decoupling strategies have 
been proved to provide perfect isolation of a system from its 
environment in the infinitely fast control limit, i.e., in the case 
of very intense and very fast control pulses \cite{viola2}. The 
efficiency of decoupling strategies in concrete situations involving 
finite strength and finite duration control pulses has been analysed 
only in the specific cases of a single qubit in a nondissipative 
environment in \cite{viola1}, and for a damped harmonic oscillator in 
a zero-temperature bath in \cite{noi}. Here we have extended these 
studies to the case of a dissipative and nonzero-temperature reservoir.
We have specialized to the case of a collective vibrational mode 
of a linear ion chain, which is used as a quantum bus in linear ion 
trap quantum computers \cite{cizo}. We have shown that the parity kick 
decoupling strategy introduced in \cite{noi} can be successfully 
applied to suppress vibrational heating, which is one important 
limitation for quantum information processing in linear ion traps 
\cite{nist}. In fact heating is suppressed as soon as the decoupling 
cycle time $T_{c}$ becomes smaller than $2\pi/\omega_{c}$, where 
$\omega_{c}$ is the bath frequency cutoff, and more importantly, the 
efficiency of this suppression is not affected by the temperature of 
the bath. The parity kick method can be applied using present 
technologies and its experimental implementation in the case of 
trapped ions would be the first example of the application of 
decoupling techniques outside the field of NMR, where the so-called 
``refocusing'' techniques \cite{nmr} are easier to use because the 
involved magnetic environment is usually very slow (see however 
Ref.~\cite{bergl} for a proof-of-principle demonstration of quantum
bang-bang control in a photon polarization qubit).

We have also shown that, differently from heating, the suppression of 
vibrational decoherence is more difficult, because in a nonzero 
temperature bath, the threshold for the decoupling cycle frequency is 
determined not only by the bath frequency cutoff, but also by the 
decoherence rate, which increases for increasing temperatures. The 
parity kick cycle frequency has to be larger than {\em both} rates and 
this makes suppression of vibrational decoherence more difficult for 
higher temperatures.

\section{Acknowledgements} This work has been partially supported by 
the European Union through the IHP program  ``QUEST''.

\end{document}